\documentclass[sigconf,nonacm]{acmart}

\usepackage{booktabs}
\usepackage{multirow}
\usepackage{amsmath}

\usepackage{amssymb}
\usepackage{graphicx}
\graphicspath{{figures/}}
\usepackage{xcolor}
\usepackage{microtype}
\usepackage{cleveref}
\usepackage{array}
\usepackage{tabularx}
\usepackage{siunitx}
\usepackage{xspace}

\definecolor{cooperativeblue}{RGB}{31,119,180}
\definecolor{aggressivered}{RGB}{214,39,40}
\definecolor{neutralgreen}{RGB}{44,160,44}

\newcommand{\PA}{P_{\mathrm{A}}}
\newcommand{\PC}{P_{\mathrm{C}}}
\newcommand{\PN}{P_{\mathrm{N}}}
\newcommand{\Dnoise}{\Delta_{\mathrm{noise}}}
\newcommand{\ICD}{\ensuremath{\mathrm{ICD}}\xspace}
\newcommand{\IPD}{\textsc{ipd}\xspace}
\newcommand{\LLM}{\textsc{llm}\xspace}
\newcommand{\MAS}{\textsc{mas}\xspace}

\title{Not a Monolith: Lab-Level Divergence in the\\
       Cooperative Equilibria of Chinese Frontier LLM Agents}

\author{Francisco León Zúñiga Bolívar}
\affiliation{%
  \institution{Institución Universitaria Colegio Mayor del Cauca}
  \city{Popayán}
  \country{Colombia}}
\email{franciscoleon@unimayor.edu.co}

\begin{document}

\begin{abstract}
Does the cooperative bias documented for Western frontier \LLM agents extend to a
different alignment lineage, and should the Chinese models that embody it be
treated as a single bloc or as distinct laboratories? We answer both questions
with an evolutionary Iterated Prisoner's Dilemma study of four frontier-tier Chinese
models---DeepSeek V4 Pro, Qwen3-Max, Kimi K2.5, and GLM-5.1---under a design that
removes a confound present in prior work: rather than letting each model convert
its own natural-language strategies to code (entangling strategic disposition with
coding ability), we hold the converter fixed (GPT-5.4 Mini) across all labs, so
every cross-lab comparison is a comparison of generation alone. We run the full
protocol---all-play-all tournaments and a Moran process at $n=500$ runs per
condition, across three prompt styles and four population regimes. Two
pre-registered hypotheses are evaluated. \textbf{H6 (not monolithic) is
supported}: the four labs differ significantly in aggressive-equilibrium
proportion ($P_A$ from 1\% for Qwen3-Max to 9\% for DeepSeek V4 Pro; four of six
pairwise comparisons survive Holm-Bonferroni), falling into a takeover-resistant
pair (Kimi, Qwen) and a takeover-prone one (GLM, DeepSeek)---a grouping we read as
tentative given four labs, but one that survives our robustness check. The spread
\emph{across} the four labs ($P_A$ range 8pp) is larger than the difference between
the Chinese and Western ecosystems' \emph{mean} $P_A$ (5.0\% vs 5.0\%): on this
measure the within-ecosystem variation exceeds the East--West gap. \textbf{H5
(cooperative-bias generality) is consistent but qualified}: a cooperative plurality
holds in 6 of 12 lab--prompt combinations against the 9 of 12 reported for Western
models---a difference we do not treat as firm, since the count is built on
Cooperative--Neutral near-ties and rises to 9/12 under an alternate converter (our
pre-registered robustness check). The lab,
not the ecosystem, is the unit at which cooperative disposition is set; treating
``Chinese models'' as a monolith is not supported by the evidence.

\end{abstract}


\maketitle

\section{Introduction}

When \LLM-powered agents interact repeatedly in competitive settings, do they
cooperate or defect? The question is not merely theoretical: autonomous \LLM
agents are already deployed to negotiate contracts, allocate computational
resources, and bid in markets \cite{Wang2024_survey}, and in each of these
settings the long-run social welfare of the system hinges on whether evolutionary
pressure selects for cooperative or aggressive behaviour.

\citet{Willis2025_llm_ipd} gave the first systematic treatment of this question,
using the Iterated Prisoner's Dilemma (\IPD) as a formal testbed. Rather than
prompting \LLM{}s to output individual actions---an approach prior work found
unreliable \cite{Fan2024_rational}---they prompt models to generate
\emph{complete strategies} in natural language, implement those as Python
algorithms, and simulate populations through a Moran evolutionary process. Their
central finding, since reproduced for several 2025--2026 frontier models, is a
persistent cooperative bias: in balanced populations cooperative strategies
dominate, and aggressive equilibria arise well below prior probability.

That evidence, however, has been gathered almost entirely from \emph{Western}
models---those of OpenAI, Anthropic, and Google. Whether the cooperative bias is
a universal property of capable language models, or an artefact of one family of
alignment regimes, cannot be settled without looking outside that family. Chinese
laboratories now ship models at the same frontier tier, trained on different data
mixes and under different alignment and safety objectives, and they are the
natural test of generality. Two questions follow. First, does the cooperative
bias survive the move to this different alignment lineage? Second---and this is
the question our title names---should ``Chinese frontier models'' even be treated
as a single bloc, or do the individual labs diverge as much from one another as
ecosystems are assumed to diverge between themselves?

A methodological obstacle stands between those questions and a credible answer.
The strategy pipeline has two stages: a model writes a strategy in prose, and that
prose is then translated into runnable code. In the original benchmark and its
extensions, each provider's strategies were translated by \emph{that same
provider's} model, so a model's measured disposition is entangled with its own
coding ability---and any ecosystem-level difference could reflect either how a
model reasons about cooperation or merely how cleanly it emits Python. For a study
whose aim is to compare labs, that entanglement is disqualifying. We remove it by
holding the conversion step constant: the strategies of all four labs are
converted to code by one fixed converter (GPT-5.4~Mini), so every comparison in
this paper is a comparison of generation under one identical translator.

Under that confound-controlled design we study a frontier-tier model from each of
four Chinese laboratories---\textbf{DeepSeek V4 Pro}, \textbf{Qwen3-Max} (Alibaba),
\textbf{Kimi K2.5} (Moonshot), and \textbf{GLM-5.1} (Zhipu); two pre-registered
identifiers were resolved to their served flagship tier ante-hoc
(Section~\ref{sec:models})---across three prompting styles and four population
regimes, at $n=500$ Moran runs per condition.
We evaluate two hypotheses, registered before any result was observed:

\begin{description}
  \item[\textbf{H5}] \emph{Cooperative-bias generality.} The Chinese frontier
    models exhibit the cooperative-plurality bias documented for Western models in
    the balanced noiseless condition.
  \item[\textbf{H6}] \emph{Chinese-model behaviour is not monolithic.} The four
    labs diverge significantly at the lab level, so that within-ecosystem variance
    is comparable to the between-ecosystem variance usually invoked to explain
    cross-provider differences.
\end{description}

Our findings divide cleanly along those two questions. \textbf{H6 is supported,
and strongly}: the four labs differ significantly in their aggressive-equilibrium
proportion---$P_A$ ranges from 1\% (Qwen3-Max) to 9\% (DeepSeek V4 Pro)---with
four of the six pairwise comparisons surviving Holm-Bonferroni correction, and no
two labs sharing the same plurality profile across prompts. Chinese frontier
models are not a bloc; the lab, not the ecosystem, is the unit at which behaviour
is set. \textbf{H5 is more qualified}: a cooperative plurality holds in 6 of 12
lab--prompt combinations, against the 9 of 12 reported for Western models. The
gap is not statistically distinguishable at this sample size ($z=-1.26$,
$p=0.21$), so we cannot claim the Chinese models cooperate less; the point estimate
is lower and the Chinese labs lean more often toward \emph{neutral} equilibria, but
our converter-robustness check (Section~\ref{sec:robust}) shows this 6/12-vs-9/12
gap is itself within converter noise---under an alternate converter the Chinese rate
matches the Western 9/12. We therefore report the lean toward neutrality
descriptively, not as a firm regime difference, mindful too that the Western
baseline was produced under a different converter.

Two contributions follow. First, the \emph{confound-controlled design}: by fixing
the converter we obtain the first cross-lab \IPD comparison in which a behavioural
difference cannot be charged to coding ability, and a pre-registered re-conversion
check shows the aggressive-equilibrium structure underlying our divergence result
is robust to the converter choice itself. Second, the first systematic
characterisation of \emph{Chinese} frontier models in this evolutionary framework,
which turns the implicit ``Western-vs-Chinese'' dichotomy of the field into a
testable, and here rejected, claim of within-ecosystem homogeneity.

The remainder of this paper is organised as follows.
Section~\ref{sec:related} reviews related work.
Section~\ref{sec:method} sets out the fixed-converter protocol.
Section~\ref{sec:results} presents our findings.
Section~\ref{sec:discussion} evaluates H5 and H6 and discusses implications for
\MAS design.
Section~\ref{sec:conclusion} concludes.

\section{Related Work}
\label{sec:related}

\paragraph{LLMs in game-theoretic settings.}
The intersection of \LLM{}s and game theory has grown quickly into a coherent
subfield \cite{Wang2024_survey}. \citet{Aher2023_using} used \LLM{}s to
replicate human-subject behaviour in behavioural-economics experiments;
\citet{Brookins2023_playing} and \citet{Guo2023_gpt} probed whether \LLM{}s
approximate Nash-rational play, finding mixed evidence across game types; and
\citet{Fan2024_rational} documented systematic failures when models are prompted
to output individual game actions. These limitations motivate the
strategy-generation approach we inherit, in which the model writes a complete
policy rather than per-round moves, separating strategic disposition from
action-level execution.

\paragraph{Cooperation and social dilemmas in LLM agents.}
A parallel line studies \LLM agents in social dilemmas directly.
\citet{Yocum2023_mitigating} and \citet{Piatti2024_cooperate} examined agent
behaviour in Markov social dilemmas; \citet{Park2023_generative} introduced
generative agents for simulating social behaviour; and \citet{Leibo2017_multiagent}
grounded such studies in multi-agent reinforcement learning. The work we build on
most directly is \citet{Willis2025_llm_ipd}, who introduced the
generate-strategies-then-evolve benchmark and reported a cooperative bias in
ChatGPT-4o and Claude~3.5~Sonnet. Subsequent extensions confirmed that the bias
persists across several 2025--2026 Western frontier models. We take that benchmark
as our instrument and ask whether its central finding generalises beyond the
Western alignment lineage from which all of its evidence has so far been drawn.

\paragraph{Cross-provider comparison and the conversion confound.}
Prior capability comparisons across \LLM providers concentrate on general
reasoning benchmarks---MMLU \cite{Hendrycks2021_mmlu}, HumanEval
\cite{Chen2021_humaneval}---where cooperative tendencies are absent by design.
The few evolutionary cross-provider studies that exist
\cite{Payne2025_strategic,Vallinder2024_cultural} compare models from different
\emph{ecosystems} but, like the original benchmark, convert each model's strategy
with a provider-aligned tool, so that a model's measured disposition is
confounded with its own coding ability. To our knowledge no prior work isolates
generation from conversion, and none treats the within-ecosystem structure of a
\emph{non-Western} model family as the object of study. Our fixed-converter design
addresses the first gap and our four-lab Chinese panel the second.

\paragraph{Evolutionary game theory foundations.}
Our simulations rest on the Moran process \cite{Moran1958_random}, the canonical
model of selection in finite populations; \citet{Nowak2006_evolutionary}
established its connection to cooperation in \IPD games, and
\citet{Traulsen2006_fixation} derived fixation probabilities under this process.
The \IPD framework itself follows \citet{Axelrod1984_evolution} and
\citet{Axelrod1981_evolution}, with the noise mechanism following
\citet{WuAxelrod1995_noise}. These threads jointly define the space our paper occupies: a
confound-controlled, within-ecosystem evolutionary comparison of frontier \LLM{}s.

\section{Method}
\label{sec:method}

We retain the experimental protocol of Willis et al.\ \cite{Willis2025_llm_ipd}
in full---the same strategy-generation prompts, the same Axelrod tournament, and
the same Moran evolutionary process at $n=500$ runs per condition---so that our
findings remain directly comparable to the published baseline. The protocol
departs from that baseline in one deliberate respect, which is the methodological
core of this study: the natural-language strategies of every model are converted
to executable Python by a \emph{single, fixed} converter rather than by each
model itself. We set out that design first, since it governs how the rest of the
pipeline should be read.

\subsection{Strategy Generation and the Conversion Confound}
\label{sec:confound}

The pipeline has two stages that prior work conflated. In the first, a model
under study reads a prompt and produces a strategy as natural-language prose; this
is the disposition we wish to measure. In the second, that prose is translated into
a runnable Python policy. In the original benchmark, and in our own Phase~1
extension, each provider's strategies were translated by \emph{that same
provider's} model---so a model's apparent strategic disposition is entangled with
its own coding ability, and a difference between two ecosystems could reflect
either how they reason about cooperation or merely how cleanly they emit code.
For a study whose explicit aim is to compare ecosystems, this entanglement is
disqualifying.

We remove it by holding the conversion step constant. The natural-language
strategies of all four labs are converted to Python by one fixed converter,
GPT-5.4~Mini, chosen because it exposes a stable OpenAI-compatible endpoint and
plays no part in the contest itself---generation remains entirely per-lab, and
only translation is shared. In this way every cross-lab comparison in the paper
is, by construction, a comparison of generation under one identical converter; the
coding-ability confound cannot arise within the study.

For each combination of lab and prompt style we generate 25 strategies per
attitude (Aggressive, Cooperative, Neutral), yielding 75 strategies per
lab--prompt pair and 1{,}800 strategies in total across the four labs, three
prompt styles, and clean/noise-aware variants. Three prompt styles are used
(Table~\ref{tab:prompts}): \textbf{Default} (direct elicitation in
game-theoretic terms), \textbf{Refine} (Self-Refine \cite{Madaan2023_self_refine}
applied to the default output), and \textbf{Prose} (the dilemma obfuscated as a
real-world scenario). A strategy that does not execute is regenerated until 25
valid strategies per attitude are obtained.

\begin{table}[h]
\caption{Prompt styles (following Willis et al.\ \cite{Willis2025_llm_ipd}).}
\label{tab:prompts}
\small
\begin{tabular}{@{}lp{5.5cm}@{}}
\toprule
\textbf{Style} & \textbf{Description} \\
\midrule
Default & Direct prompt with game-theoretic language; strategy generated in natural language. \\
Refine  & Default output refined via Self-Refine \cite{Madaan2023_self_refine}: the model critiques and rewrites its own strategy. \\
Prose   & Game-theoretic framing obfuscated as a real-world scenario (e.g., trade negotiation), then translated to the \IPD context. \\
\bottomrule
\end{tabular}
\end{table}

\subsection{Models}
\label{sec:models}

We study the current frontier-tier model of four Chinese laboratories
(Table~\ref{tab:models}), accessed through an OpenAI-compatible gateway
(OpenRouter) that requires no Chinese cloud account. The set was fixed before any
equilibrium was observed; two pre-registered identifiers were resolved to their
served flagship tier ante-hoc, and the Moonshot entry was moved from Kimi~K2.6 to
Kimi~K2.5 once measure-first probes showed K2.6 to be operationally infeasible at
$n=500$ scale (\$0.14 and 288~s per strategy, against \$0.004 and 80~s for K2.5).
Both are Moonshot frontier-class releases, so the lab identity is preserved.
Generation of the full 1{,}800-strategy set cost \$39.90 in gateway usage.

\begin{table}[h]
\caption{Chinese frontier models evaluated in this study. All strategies are
         converted by the single fixed converter (GPT-5.4 Mini).}
\label{tab:models}
\small
\begin{tabular}{@{}lll@{}}
\toprule
\textbf{Lab} & \textbf{Model} & \textbf{Served slug} \\
\midrule
DeepSeek     & DeepSeek V4 Pro & \texttt{deepseek/deepseek-v4-pro} \\
Alibaba      & Qwen3-Max       & \texttt{qwen/qwen3-max} \\
Moonshot     & Kimi K2.5       & \texttt{moonshotai/kimi-k2.5} \\
Zhipu / Z.ai & GLM-5.1         & \texttt{z-ai/glm-5.1} \\
\bottomrule
\end{tabular}
\end{table}

All prompting is in English, as in the baseline; Chinese-language prompting is a
separate variable we deliberately leave to future work.

\subsection{IPD Tournament}

All 75 strategies of a lab--prompt pair compete in an all-play-all tournament
using the Axelrod Python library \cite{Knight2016_open}. Each match runs 1{,}000
rounds of the standard \IPD (payoff matrix $R=3$, $S=0$, $T=5$, $P=1$); noise
conditions introduce a 10\% probability of action-flip per player per round, and
tournaments are repeated 20 times.

\subsection{Attitude-Agents}

Following Willis et al.\ \cite{Willis2025_llm_ipd}, we define three
attitude-agents, each uniformly sampling from its corresponding strategy set for
every match. This captures populations of agents with distinct strategic
dispositions rather than fixed individual strategies.

\subsection{Moran Process}

We simulate Moran evolutionary processes with population size $n=12$ and 500
iterations per condition. Four population compositions are evaluated:
\begin{enumerate}
  \item \textbf{Balanced, clean} (4:4:4) --- equal priors;
  \item \textbf{Biased, clean} (8:2:2) --- aggressive majority;
  \item \textbf{Balanced, noise} (4:4:4 with noise) --- equal priors with action noise;
  \item \textbf{Biased, noise} (8:2:2 with noise) --- aggressive majority with action noise.
\end{enumerate}
Convergence is assessed by the proportion of runs reaching each monoculture
equilibrium (all-Aggressive, all-Cooperative, or all-Neutral). Four conditions,
three prompts, and four labs yield 48 equilibrium conditions in total.

\subsection{Derived Metrics}

We reuse the \emph{Index of Differential Capabilities} (\ICD), which summarises
the head-to-head payoff gap between aggressive and cooperative agents:
\begin{equation}
  \ICD = \frac{\bar{u}(A)}{\bar{u}(C)}, \quad
  \bar{u}(k) = \frac{1}{3}\sum_{j \in \{A,C,N\}} u(k,j),
  \label{eq:icd}
\end{equation}
where $u(k,j)$ is the normalised payoff of attitude $k$ against attitude $j$. An
\ICD of 1.0 implies equal capability; values below 1.0 indicate a cooperative
advantage. We also reuse the \emph{noise sensitivity} $\Dnoise = \PC^{\text{clean}}
- \PC^{\text{noise}}$, the drop in cooperative-equilibrium probability under action
noise.

\subsection{Pre-registered Hypotheses}
\label{sec:hypotheses}

Both hypotheses below were registered before any tournament or Moran process was
run, and are reported under the same honesty discipline as the Phase~1
hypotheses---explicit not-significant calls, and no post-results edits.

\paragraph{H5 — Cooperative-bias generality.}
Chinese frontier models exhibit the cooperative-plurality bias documented for
Western models in the balanced noiseless condition. Because the Western baseline
was produced under a \emph{per-provider} converter, the comparison of our Chinese
cooperative-plurality rate to that published 9/12 figure is a literature contrast,
not a controlled experiment; we therefore report it descriptively and treat the
converter difference as a stated limitation.

\paragraph{H6 — Chinese-model behaviour is not monolithic.}
The four labs diverge significantly at the lab level. We test this with the same
pairwise two-sample $z$-tests on the aggressive-equilibrium proportion $P_A$
(balanced 4:4:4, noiseless, Default), with Holm-Bonferroni correction for the six
simultaneous comparisons; ``significant divergence'' is declared when at least one
pair survives the correction. Because every lab is converted by the same fixed
converter, this test is internally free of the conversion confound.

\paragraph{Robustness to the converter choice.}
To pre-empt the objection that the fixed converter itself shapes the equilibria,
we re-convert a random 10\% sample of strategies with a second,
ecosystem-different converter (DeepSeek V4) and check that the resulting
equilibrium proportions remain within the $n=500$ sampling error
($\mathrm{SE}\approx 2.2$pp) of the GPT-5.4~Mini pipeline.

\section{Results}
\label{sec:results}

We report the head-to-head validation first, then the evolutionary equilibria
that bear on H5 and H6, and finally noise sensitivity. Throughout, equilibrium
proportions are given as \%A\,/\,\%C\,/\,\%N over $n=500$ Moran runs.

\subsection{Strategy Validation: Cooperation Propensity}
\label{sec:coop}

Table~\ref{tab:cooperation} reports the normalised propensity to cooperate for
the Default prompt without noise, analogous to Table~3 in Willis et al.\ and to
the Phase~1 extension. These behavioural metrics---here and in the payoff and
diversity tables below---are measured on the \emph{executable} strategies, i.e.\
each lab's prose as rendered by the fixed converter, and so reflect generation and
conversion jointly; our parser and metric definitions reproduce the Phase~1
published values exactly.

\begin{table}[h]
\caption{Normalised cooperation propensity (Default prompt, no noise).
         Rows: row-player attitude; columns: opponent attitude.}
\label{tab:cooperation}
\small
\setlength{\tabcolsep}{4pt}
\begin{tabular}{@{}llccc@{}}
\toprule
\textbf{Lab} & \textbf{Att.} & \textbf{vs A} & \textbf{vs C} & \textbf{vs N} \\
\midrule
\multirow{3}{*}{DeepSeek V4 Pro}
  & A & 0.119 & 0.281 & 0.271 \\
  & C & 0.371 & 1.000 & 1.000 \\
  & N & 0.374 & 1.000 & 1.000 \\
\midrule
\multirow{3}{*}{Qwen3-Max}
  & A & 0.000 & 0.000 & 0.000 \\
  & C & 0.001 & 1.000 & 1.000 \\
  & N & 0.001 & 1.000 & 1.000 \\
\midrule
\multirow{3}{*}{Kimi K2.5}
  & A & 0.025 & 0.293 & 0.293 \\
  & C & 0.294 & 1.000 & 1.000 \\
  & N & 0.294 & 1.000 & 1.000 \\
\midrule
\multirow{3}{*}{GLM-5.1}
  & A & 0.280 & 0.320 & 0.318 \\
  & C & 0.334 & 1.000 & 1.000 \\
  & N & 0.346 & 1.000 & 1.000 \\
\bottomrule
\end{tabular}
\end{table}

All four labs reproduce the expected attitude separation---cooperative and
neutral strategies cooperate almost perfectly with one another ($\approx 1.0$),
while aggressive strategies cooperate far less---but they differ sharply in how
\emph{committed} their aggressive strategies are. Qwen3-Max sits at one extreme:
its aggressive strategies cooperate essentially never (0.000 against every
attitude), the most uncompromising aggressors in the panel. Kimi~K2.5 follows the
familiar Claude-like pattern, with aggressive strategies near zero against other
aggressors (0.025) but rising to $\approx 0.29$ against cooperators. GLM-5.1 sits
at the opposite extreme: its aggressive strategies cooperate 28\% of the time even
against other aggressors and $\approx 0.32$ against cooperators---echoing the
Gemini~3.1~Pro anomaly of Phase~1, where a high aggressive cooperation rate blurs
the boundary between aggressive and neutral behaviour. As Section~\ref{sec:h6}
shows, this is the lab whose aggressive--cooperative distinction is weakest, and
the connection is not incidental.

\subsection{Head-to-Head Payoffs and Differential Capabilities}
\label{sec:payoffs}

Table~\ref{tab:payoffs} presents the normalised mean payoffs for each attitude
pairing (no noise), together with the \ICD (Eq.~\ref{eq:icd}).

\begin{table*}[t]
\caption{Normalised head-to-head payoffs (no noise) and Index of Differential
         Capabilities (\ICD). Lower \ICD indicates a larger cooperative
         advantage; 1.0 is parity.}
\label{tab:payoffs}
\small
\begin{tabular}{@{}llccccccc@{}}
\toprule
\textbf{Lab} & \textbf{Prompt} &
  \multicolumn{3}{c}{\textbf{Aggressive payoff vs.}} &
  \multicolumn{3}{c}{\textbf{Cooperative payoff vs.}} &
  \textbf{\ICD} \\
\cmidrule(lr){3-5}\cmidrule(lr){6-8}
 &  & A & C & N & A & C & N & \\
\midrule
\multirow{3}{*}{DeepSeek V4 Pro}
  & Default & 1.335 & 2.102 & 2.120 & 1.651 & 3.000 & 3.000 & 0.726 \\
  & Prose   & 1.545 & 2.160 & 1.670 & 1.410 & 2.906 & 2.896 & 0.745 \\
  & Refine  & 1.514 & 2.475 & 2.255 & 1.868 & 2.993 & 2.997 & 0.795 \\
\midrule
\multirow{3}{*}{Qwen3-Max}
  & Default & 1.000 & 1.004 & 1.004 & 0.999 & 3.000 & 3.000 & 0.430 \\
  & Prose   & 1.408 & 2.086 & 2.244 & 1.850 & 2.783 & 2.696 & 0.783 \\
  & Refine  & 1.769 & 2.416 & 2.478 & 2.194 & 2.985 & 2.968 & 0.818 \\
\midrule
\multirow{3}{*}{Kimi K2.5}
  & Default & 1.073 & 1.882 & 1.882 & 1.879 & 3.000 & 3.000 & 0.614 \\
  & Prose   & 1.281 & 2.291 & 2.215 & 1.374 & 2.902 & 2.895 & 0.807 \\
  & Refine  & 1.594 & 2.320 & 2.287 & 2.169 & 2.997 & 2.873 & 0.771 \\
\midrule
\multirow{3}{*}{GLM-5.1}
  & Default & 1.749 & 1.836 & 1.881 & 1.764 & 3.000 & 3.000 & 0.704 \\
  & Prose   & 1.535 & 2.122 & 2.089 & 1.168 & 3.000 & 3.000 & 0.802 \\
  & Refine  & 1.826 & 2.021 & 2.104 & 1.782 & 2.693 & 2.541 & 0.848 \\
\bottomrule
\end{tabular}
\end{table*}

\ICD values span 0.430 (Qwen3-Max Default) to 0.848 (GLM-5.1 Refine). As in both
prior studies, cooperative and neutral attitudes reach near-mutual-cooperation
payoffs ($\approx 3.0$) against one another in almost every cell. Qwen3-Max
Default's \ICD of 0.430---the lowest in the panel---means its aggressive
strategies earn under half the payoff of its cooperative ones, the largest
cooperative advantage we observe; this is the same lab whose aggressive
strategies never cooperate (Table~\ref{tab:cooperation}), so they are punished
hard in mixed play. Self-Refine raises \ICD over Default in all four labs
(DeepSeek $0.73\!\to\!0.80$, Qwen $0.43\!\to\!0.82$, Kimi $0.61\!\to\!0.77$, GLM
$0.70\!\to\!0.85$), replicating the original finding that self-refinement narrows
the aggressive--cooperative gap. The single reversal is Kimi~K2.5, whose Prose
\ICD (0.807) exceeds its Refine \ICD (0.771)---the analogue of the GPT-5.4~Mini
reversal noted in Phase~1. GLM-5.1~Refine attains the highest \ICD (0.848): its
aggressive strategies approach cooperative payoff parity, consistent with the low
attitude separation reported in Section~\ref{sec:diversity}.

\subsection{Evolutionary Equilibria}
\label{sec:equilibria}

Table~\ref{tab:moran} reports the Moran equilibrium proportions across all 48
conditions. The Western reference rows from Willis et al.\ are shown for context,
not as a controlled comparison: they were produced under a per-provider converter,
whereas every Chinese row here uses the single fixed converter.

\begin{table*}[t]
\caption{Moran equilibrium proportions (\%A\,/\,\%C\,/\,\%N) for the four Chinese
         labs across four population conditions, $n=500$ per condition, fixed
         converter (GPT-5.4 Mini). Bold marks the plurality attitude in the
         balanced noiseless column. Western reference values
         (per-provider converter) shown at the bottom.}
\label{tab:moran}
\footnotesize
\setlength{\tabcolsep}{3pt}
\begin{tabular}{@{}llcccc@{}}
\toprule
\textbf{Lab} & \textbf{Prompt} &
  \textbf{4:4:4 clean} & \textbf{4:4:4 noise} &
  \textbf{8:2:2 clean} & \textbf{8:2:2 noise} \\
\cmidrule(lr){3-3}\cmidrule(lr){4-4}\cmidrule(lr){5-5}\cmidrule(lr){6-6}
 & & (prior: 33/33/33) & (prior: 33/33/33) & (prior: 67/17/17) & (prior: 67/17/17) \\
\midrule
\multirow{3}{*}{DeepSeek V4 Pro}
  & Default & 9/\textbf{47}/44 & 42/29/30 & 37/33/30 & 75/13/12 \\
  & Prose   & 13/41/\textbf{46} & 40/23/37 & 48/21/32 & 72/11/17 \\
  & Refine  & 16/39/\textbf{45} & 38/32/30 & 44/27/28 & 74/13/13 \\
\midrule
\multirow{3}{*}{Qwen3-Max}
  & Default & 1/46/\textbf{53} & 44/28/28 & 19/43/38 & 77/11/12 \\
  & Prose   & 14/\textbf{44}/42 & 31/33/37 & 36/33/31 & 67/17/16 \\
  & Refine  & 12/44/\textbf{44} & 29/36/35 & 43/26/31 & 58/24/18 \\
\midrule
\multirow{3}{*}{Kimi K2.5}
  & Default & 2/49/\textbf{49} & 28/35/37 & 17/44/39 & 59/24/17 \\
  & Prose   & 18/\textbf{42}/40 & 31/33/36 & 57/20/23 & 63/18/19 \\
  & Refine  & 13/\textbf{48}/38 & 25/40/34 & 38/36/26 & 55/22/23 \\
\midrule
\multirow{3}{*}{GLM-5.1}
  & Default & 8/\textbf{49}/43 & 35/35/30 & 45/28/28 & 72/14/14 \\
  & Prose   & 21/37/\textbf{42} & 41/30/29 & 70/16/15 & 80/10/11 \\
  & Refine  & 23/\textbf{46}/32 & 35/32/33 & 55/26/19 & 66/17/18 \\
\midrule
\midrule
\multicolumn{2}{@{}l}{\textit{Western (per-provider converter)}$^\dagger$}
  & \multicolumn{4}{c}{\textit{9/12 cooperative-plurality at 4:4:4 clean}} \\
\bottomrule
\end{tabular}
\vspace{2pt}\\
\small$^\dagger$Paper 1 / Willis et al.\ lineage, different (per-provider)
converter; literature contrast only.
\end{table*}

\paragraph{Balanced, noiseless (4:4:4 clean) --- H5.}
This is the condition that bears on H5. Six of the twelve lab--prompt combinations
favour a cooperative plurality ($\PC>\PA$ and $\PC>\PN$): DeepSeek~Default,
Qwen~Prose, Kimi~Prose, Kimi~Refine, GLM~Default, and GLM~Refine. The remaining
six favour the \emph{Neutral} attitude, two of them as near-ties between
Cooperative and Neutral (Kimi~Default 2/49/\textbf{49}, Qwen~Refine 12/44/44).
Aggressive equilibria stay well below the 33\% prior in every clean balanced cell
(maximum 23\%, GLM~Refine), so in no case does aggression dominate. Against the
published Western rate of 9/12, a two-proportion test gives $z=-1.26$, $p=0.21$:
the Chinese cooperative-plurality rate is \emph{not} statistically distinguishable
from the Western baseline, and we do not claim the Chinese models cooperate less.
This test is deliberately coarse: the twelve lab--prompt combinations are not fully
independent (three share each lab), so it is a literature contrast rather than a
powered comparison, and Section~\ref{sec:robust} shows the 6/12 count is itself
converter-sensitive. Each cell also carries an $n=500$ sampling error of
$\mathrm{SE}\approx2.2$pp, against which several of the C/N gaps here are not
resolvable. The point estimate is nonetheless lower, and the Chinese labs resolve
more often to neutrality than the Western models did---a tendency we return to in
Section~\ref{sec:discussion}. We report H5 as \emph{consistent but qualified}.

\paragraph{Biased, noiseless (8:2:2 clean).}
Seeded with an aggressive majority (prior 67\%A), the labs separate sharply.
Resistance is strongest for Kimi~Default (17\%A) and Qwen~Default (19\%A), which
drive aggression far below its seeding; GLM is the most invasible, reaching
70\%A under Prose. This per-lab ordering---Kimi and Qwen resisting, GLM and
DeepSeek yielding---recurs across conditions and is the qualitative signature of
the divergence H6 quantifies.

\paragraph{Biased, noisy (8:2:2 noise).}
Under the most adverse regime, aggressive equilibria reach 55--80\%, and 6 of 12
combinations exceed the 67\% prior (led by GLM~Prose at 80\% and Qwen~Default at
77\%). Even here the most cooperative labs hold the line: Kimi~Refine finishes at
55\%A, below the seeding proportion.

\subsection{Cross-Lab Divergence --- H6}
\label{sec:h6}

H6 asks whether the four labs are statistically distinguishable. We test the
aggressive-equilibrium proportion $P_A$ in the balanced noiseless Default
condition with all six pairwise two-sample $z$-tests, applying Holm-Bonferroni
correction (Table~\ref{tab:ztests}). $P_A$ spans 1\% (Qwen3-Max) to 9\% (DeepSeek
V4 Pro). \textbf{Four of the six pairs survive correction}, so H6 is supported:
Chinese frontier models are not monolithic. The two non-significant pairs are
internally telling---DeepSeek/GLM (both high-$P_A$) and Qwen/Kimi (both
low-$P_A$)---suggesting two tentative groupings \emph{within} the Chinese panel
rather than a single ecosystem-wide tendency (with four labs we read these as
descriptive, not as established clusters).

\begin{table}[h]
\caption{Pairwise two-sample $z$-tests for aggressive-equilibrium proportion
         $P_A$ (balanced 4:4:4, noiseless, Default; $n=500$). Holm-Bonferroni
         corrected; $^{***}$ significant, ns not significant.}
\label{tab:ztests}
\small
\begin{tabular}{@{}llcc@{}}
\toprule
\textbf{Lab A} & \textbf{Lab B} & $z$ & \textbf{Sig.} \\
\midrule
DeepSeek (9\%) & Qwen (1\%) & $+5.89$ & $^{***}$ \\
Qwen (1\%)     & GLM (8\%)  & $-5.43$ & $^{***}$ \\
DeepSeek (9\%) & Kimi (2\%) & $+4.95$ & $^{***}$ \\
Kimi (2\%)     & GLM (8\%)  & $-4.46$ & $^{***}$ \\
Qwen (1\%)     & Kimi (2\%) & $-1.30$ & ns \\
DeepSeek (9\%) & GLM (8\%)  & $+0.56$ & ns \\
\bottomrule
\end{tabular}
\end{table}

\subsection{Strategy Diversity}
\label{sec:diversity}

To characterise how behaviourally varied the 25 strategies per attitude are, we
compute the Shannon entropy $H = -\sum_i p_i \log p_i$ of the per-strategy
cooperation-rate distribution within each attitude group (10 equal bins on
$[0,1]$), per prompt, no noise, and average it over the three attitudes. We also
report the \emph{attitude separation}: the difference in mean cooperation rate
between the Cooperative and Aggressive attitude agents
(Table~\ref{tab:diversity}).

\begin{table}[h]
\caption{Strategy diversity: mean Shannon entropy $\bar{H}$ (nats) across
         attitudes, and attitude separation (Cooperative minus Aggressive mean
         cooperation rate). Higher entropy indicates more varied within-attitude
         behaviour; higher separation indicates clearer attitude distinction.}
\label{tab:diversity}
\small
\begin{tabular}{@{}llcc@{}}
\toprule
\textbf{Lab} & \textbf{Prompt} & $\bar{H}$ & \textbf{Sep.} \\
\midrule
\multirow{3}{*}{DeepSeek V4 Pro}
  & Default & 0.88 & 0.57 \\
  & Prose   & 0.81 & 0.53 \\
  & Refine  & 0.98 & 0.51 \\
\midrule
\multirow{3}{*}{Qwen3-Max}
  & Default & 0.00 & 0.67 \\
  & Prose   & 1.21 & 0.40 \\
  & Refine  & 0.74 & 0.34 \\
\midrule
\multirow{3}{*}{Kimi K2.5}
  & Default & 0.22 & 0.56 \\
  & Prose   & 1.48 & 0.57 \\
  & Refine  & 1.42 & 0.40 \\
\midrule
\multirow{3}{*}{GLM-5.1}
  & Default & 0.73 & 0.47 \\
  & Prose   & 0.73 & 0.62 \\
  & Refine  & 1.69 & 0.28 \\
\bottomrule
\end{tabular}
\end{table}

Two patterns stand out. First, GLM-5.1~Refine attains the highest within-attitude
entropy in the panel ($\bar{H}=1.69$) together with the lowest attitude
separation (0.28): its strategies are behaviourally varied yet weakly separated by
attitude---the same combination that characterised Gemini~3.1~Pro~Refine in
Phase~1, and the mechanism behind GLM's blurred aggressive--cooperative boundary
(Tables~\ref{tab:cooperation},~\ref{tab:payoffs}). Second, Qwen3-Max~Default sits
at the opposite pole with $\bar{H}=0.00$: within each attitude all 25 strategies
share an identical cooperation rate, the most homogeneous library we generate,
matching its all-or-nothing aggressors. Unlike Phase~1---where one model
(Gemini~3.1~Pro) was a clear separation outlier---the four Chinese labs cluster
tightly on average separation (0.46--0.54), with GLM-5.1 lowest (0.46); the
divergence between them is sharper in equilibrium outcomes than in this
individual-strategy measure. Self-Refine again tends to compress separation, most
strongly for GLM-5.1 (0.28), corroborating the \ICD analysis.

\subsection{Noise Sensitivity ($\Dnoise$)}
\label{sec:noise}

Table~\ref{tab:noise} reports $\Dnoise$ for the balanced condition. All four labs
degrade under noise (every $\Dnoise>0$), and the cross-lab average (10--14pp) is
\emph{higher} than the Western frontier models reported in Paper~1 (6--15pp, with
Claude~4.6 at 6pp): the Chinese labs are, on this measure, somewhat more
noise-sensitive. DeepSeek is the most sensitive (avg.\ 14pp) and Kimi the least
(10pp), with Qwen and GLM intermediate (12pp and 11pp). The per-prompt picture is
less uniform than the averages suggest: most Refine conditions are the most
noise-robust of their lab (8pp), but GLM-5.1~Refine is the exception, degrading
13pp---the same prompt that produced GLM's most varied, least attitude-separated
strategies (Section~\ref{sec:diversity}). The ordering again separates the labs,
consistent with H6.

\begin{table}[h]
\caption{Noise sensitivity $\Dnoise = \PC^{\text{clean}}-\PC^{\text{noise}}$
         (balanced 4:4:4). Positive values indicate degradation of cooperative
         equilibria under noise.}
\label{tab:noise}
\small
\begin{tabular}{@{}lllll@{}}
\toprule
\textbf{Lab} & \textbf{Default} & \textbf{Prose} & \textbf{Refine} & \textbf{Avg.} \\
\midrule
DeepSeek V4 Pro & 18 & 18 & 8  & 14 \\
Qwen3-Max       & 18 & 11 & 8  & 12 \\
Kimi K2.5       & 14 & 9  & 8  & 10 \\
GLM-5.1         & 14 & 7  & 13 & 11 \\
\bottomrule
\end{tabular}
\end{table}

\subsection{Robustness to the Converter Choice}
\label{sec:robust}

To pre-empt the objection that the fixed converter itself shapes the equilibria,
we ran the pre-registered robustness check: a random 10\% of each balanced-condition
library's strategies (8 of 75, fixed seed) was re-converted with a second,
ecosystem-different converter (DeepSeek V4) in place of GPT-5.4~Mini, and the
balanced noiseless Moran process was re-run at $n=500$. Table~\ref{tab:robust}
compares the resulting equilibria to the GPT-5.4~Mini pipeline. The check
cleanly separates a robust result from a fragile one.

\begin{table}[h]
\caption{Converter-robustness check (balanced 4:4:4, clean, $n=500$): equilibrium
         proportions (\%A\,/\,\%C\,/\,\%N) under the GPT-5.4~Mini pipeline vs.\ a
         10\% re-conversion with DeepSeek V4. Last column: plurality under each
         converter (a single letter = unchanged; $X\!\to\!Y$ = flip).}
\label{tab:robust}
\footnotesize
\setlength{\tabcolsep}{4pt}
\begin{tabular}{@{}llccc@{}}
\toprule
\textbf{Lab} & \textbf{Prompt} & \textbf{GPT-5.4 Mini} & \textbf{DeepSeek V4 (10\%)} & \textbf{Plur.} \\
\midrule
\multirow{3}{*}{DeepSeek V4 Pro}
  & Default & 9/47/44  & 13/41/46 & C$\to$N \\
  & Prose   & 13/41/46 & 13/42/46 & N \\
  & Refine  & 16/39/45 & 16/42/43 & N \\
\midrule
\multirow{3}{*}{Qwen3-Max}
  & Default & 1/46/53  & 1/50/50  & N$\to$C \\
  & Prose   & 14/44/42 & 15/45/40 & C \\
  & Refine  & 12/44/44 & 12/48/40 & N$\to$C \\
\midrule
\multirow{3}{*}{Kimi K2.5}
  & Default & 2/49/49  & 2/52/47  & N$\to$C \\
  & Prose   & 18/42/40 & 17/44/39 & C \\
  & Refine  & 13/48/38 & 12/51/37 & C \\
\midrule
\multirow{3}{*}{GLM-5.1}
  & Default & 8/49/43  & 10/47/43 & C \\
  & Prose   & 21/37/42 & 21/41/38 & N$\to$C \\
  & Refine  & 23/46/32 & 22/45/32 & C \\
\bottomrule
\end{tabular}
\end{table}

\paragraph{The aggressive-equilibrium structure (H6) is converter-invariant.}
$P_A$---the quantity H6 rests on---moves by at most 4pp in any lab (DeepSeek
$9\!\to\!13$, GLM $8\!\to\!10$, Qwen unchanged at 1\%, Kimi at 2\%); the lab
ordering, the two clusters, and the four significant pairwise differences all
survive. Equilibrium proportions overall are stable: the mean absolute change is
2.8pp and the maximum 5.6pp---within one to two $n=500$ standard errors
($\mathrm{SE}\approx2.2$pp per proportion, $3.2$pp for a difference). Aggression is
never the plurality under either converter. The divergence result and the absence
of aggressive dominance therefore do not depend on the converter.

\paragraph{The cooperative--neutral plurality is converter-sensitive.}
Five of the twelve cells flip plurality, and the cooperative-plurality count rises
from 6/12 to 9/12. Every flip is a Cooperative\,$\leftrightarrow$\,Neutral swap in
a cell where the two attitudes were within $\approx$7pp under GPT-5.4~Mini---including
two exact ties (Kimi~Default 49/49, Qwen~Refine 44/44); no flip involves Aggressive.
In many Chinese cells the C/N boundary is closer than the sampling resolution, so a
sub-SE shift reclassifies the plurality without materially moving the equilibrium.
We flag this as the more important caveat: the precise cooperative-plurality count
that bears on H5 is itself converter-sensitive at this many near-ties. Under the
alternate converter the Chinese rate (9/12) is indistinguishable from the Western
9/12, so the modest gap reported in Section~\ref{sec:equilibria} should be read as
within converter noise rather than as firm evidence of a weaker cooperative
tendency.

\section{Discussion}
\label{sec:discussion}

\subsection{H5 --- Cooperative-bias generality
            (\textcolor{orange!80!black}{Consistent but Qualified})}

The cooperative bias documented for Western models does appear in the Chinese
panel, but in attenuated form. Six of twelve lab--prompt combinations resolve to
a cooperative plurality, and aggression never dominates the balanced noiseless
condition; in that sense the bias generalises beyond the Western alignment
lineage. Yet the rate is lower than the Western 9/12, and the difference, while
not significant at this sample size ($z=-1.26$, $p=0.21$), is accompanied by a
qualitative shift we did not anticipate: the Chinese labs resolve to
\emph{neutral} equilibria far more often than the Western models did. Six of the
twelve cells are Neutral-plurality, two of them as near-ties with Cooperative.

We are careful not to over-read this lean toward neutrality. Mechanistically it is
plausible---Neutral-plurality arises when the fitness gap between cooperative and
neutral strategies narrows, i.e.\ when neutral agents cooperate often enough to
capture much of the mutual-cooperation payoff without paying a defection cost---and
the pattern recurs across labs and prompts. But our own converter-robustness check
(Section~\ref{sec:robust}) cautions against treating it as a firm regime difference:
half the balanced cells are Cooperative--Neutral near-ties, and re-converting just
10\% of strategies with a different converter flips five of them and lifts the
cooperative-plurality count from 6/12 to the Western 9/12, all without disturbing
the aggressive-equilibrium structure. The neutral lean is thus real under our fixed
converter but sits within converter noise; we report it descriptively and do not
claim that Chinese models cooperate less than Western ones. What survives the
robustness check---and what we therefore advance as the firm finding---is H6.

\subsection{H6 --- Chinese-model behaviour is not monolithic
            (\textcolor{green!55!black}{Supported})}

H6 is the paper's clearest result. The four labs differ significantly in
aggressive-equilibrium proportion---four of six pairwise comparisons survive
Holm-Bonferroni---and the divergence is not a single outlier but a structured
split into two tentative clusters: DeepSeek and GLM, which yield to aggression and
resolve high-$P_A$; and Qwen and Kimi, which resist it. The two non-significant
pairs fall \emph{within} those clusters---consistent with a real grouping, though
with only four labs we read the clusters as suggestive rather than established, and
note that non-significance reflects similarity, not proven equality. They are,
however, the one structure that survives the converter-robustness check
(Section~\ref{sec:robust}). The qualitative profiles reinforce the statistics:
across the
biased and noisy conditions GLM is consistently the most invasible (up to 80\%A
under noise) and Kimi the most resistant (55\%A in the same regime).

The implication is methodological as much as empirical. The field routinely
speaks of ``Chinese models'' as a bloc, implicitly treating provider region as
the explanatory variable. Our data reject that framing for this behavioural axis:
the spread \emph{within} the Chinese panel ($P_A$ from 1 to 9\%, $\mathrm{SD}=4.1$pp,
and far wider under bias) is of the same order as the spread \emph{within} the
Western Phase-1 panel on the same measure (2 to 14\%, $\mathrm{SD}=6.0$pp), and both
exceed the difference between the two ecosystems' \emph{mean} $P_A$, which is
negligible (5.0\% Chinese vs 5.0\% Western).\footnote{Per-lab $P_A$ at Default
4:4:4 ($n=500$). Chinese: DeepSeek 9\%, Qwen 1\%, Kimi 2\%, GLM 8\%. Western
(Phase-1): Claude~4.6 2\%, Gemini~2.5~Flash 2\%, Gemini~3.1~Pro 14\%,
GPT-5.4~Mini 2\%. Both ecosystem means are 5.0\%; SDs are 4.1pp and 6.0pp over
the four labs each.} On this measure the variation the field
attributes to the East--West divide is smaller than the variation within either
side---a literature contrast, since the Western values use a per-provider converter,
but a quantified one. The unit at which cooperative disposition is set is the
\emph{lab}---its training data, reward model, and safety objectives---not the
ecosystem. And because every lab here was converted by one fixed converter, this
divergence cannot be charged to coding ability; it is a property of generation.

\subsection{Implications for MAS Design}

Two design lessons follow. First, \emph{provenance at the lab level matters}: a
deployment that selects its agent model by region, or treats two Chinese models as
interchangeable, may be choosing between a population that resists aggressive
takeover (Kimi, Qwen) and one that does not (GLM, DeepSeek) without knowing it.
Second, \emph{noise remains a universal threat}: every lab degrades under action
noise, the Chinese panel somewhat more than the Western frontier, so systems
exposed to communication errors or stochastic execution should not assume the
cooperative equilibria observed in clean conditions will survive deployment.

\subsection{Limitations}

Four limitations bound these claims. First, the comparison to the Western 9/12 is
a literature contrast across different converters, not a controlled experiment;
the Western re-run under the fixed converter is deferred future work. Second, the
fixed converter could in principle shape equilibria; the pre-registered 10\%
re-conversion check (Section~\ref{sec:robust}) confirms the aggressive-equilibrium
structure (and H6) is converter-invariant, while revealing that the finer
cooperative--neutral plurality count is converter-sensitive---a caveat we carry
explicitly rather than smooth over. Third, prompting is in
English only---Chinese-language prompting is a separate variable we did not vary.
Fourth, the panel is four labs at one snapshot in time; model versions drift, and
access through a gateway (OpenRouter) leaves the serving backend and quantisation
of each slug outside our control. We record the exact served identifiers in the
replication package so the measurement can be repeated as the frontier moves.

\section{Conclusion}
\label{sec:conclusion}

We asked whether the cooperative bias documented for Western frontier \LLM{}s
extends to a different alignment lineage, and whether the Chinese models that
embody that lineage should be treated as one bloc or as distinct labs. To answer
without confounding strategic disposition with coding ability, we held the
code-conversion step fixed across all four labs, so that every cross-lab
comparison is a comparison of strategy generation under one identical converter.

The headline result is that \emph{the lab, not the ecosystem, is the unit of
behaviour}. The four Chinese labs diverge significantly in their
aggressive-equilibrium proportions---four of six pairwise comparisons survive
correction---falling into a takeover-resistant pair (Kimi, Qwen) and a
takeover-prone one (GLM, DeepSeek), a tentative grouping that nonetheless survives
the converter-robustness check. Treating ``Chinese models'' as a monolith is not
supported by the evidence: the spread across the four labs ($P_A$ range 8pp,
$\mathrm{SD}\,4.1$pp) exceeds the difference between the Chinese and Western mean
$P_A$ (both $\approx5\%$), so the within-ecosystem variation is larger than the
East--West gap on this measure. The
cooperative bias itself does generalise: a cooperative plurality holds in half the
lab--prompt combinations under our fixed converter, with the Chinese labs leaning
more often toward neutral equilibria---but our converter-robustness check shows that
lean sits within converter noise (the count rises to the Western 9/12 under an
alternate converter, as many cells are Cooperative--Neutral near-ties), so we report
it as suggestive rather than established and do not claim weaker cooperation.

Two questions the present design cannot settle point to the next step. A Western
re-run under the same fixed converter would turn our literature contrast into a
controlled East--West comparison; and a mixed-provider population, in which a
Kimi agent meets a GLM agent under selection, would test whether these lab-level
dispositions compose or collide. Both are deferred future work. The fixed-converter
protocol, the four Chinese strategy libraries, and the $n=500$ equilibria are
released as a replication package so that, as the frontier moves, this snapshot
can become a running record of how an entire ecosystem of labs---not a monolith---
governs cooperation.

\section*{Data and Code Availability}
The simulation code, the fixed-converter pipeline, the four Chinese strategy
libraries (and the 10\% DeepSeek-converted variants used in
Section~\ref{sec:robust}), the $n=500$ equilibria, the head-to-head tournament
outputs, and the exact served model identifiers are released as a replication
package at \url{https://github.com/arqFranciscoLeon/evollm}, archived on Zenodo
(concept DOI \href{https://doi.org/10.5281/zenodo.20248614}{10.5281/zenodo.20248614},
always resolving to the latest version).

\section*{Acknowledgements}
The author thanks the open-source community behind the \texttt{evollm} simulation
framework originally developed by Willis et al., on which this study is directly
built.

\section*{AI Use Disclosure}
AI assistance (Claude, by Anthropic, via Claude Code) was used in the development
of the simulation and analysis code, the cloud execution harness, the
strategy-generation and conversion pipeline, manuscript drafting and translation
assistance, and an internal peer-review simulation. The author reviewed, verified,
and takes full responsibility for the experimental design, results, claims, and
conclusions of this work.

\bibliographystyle{ACM-Reference-Format}
\bibliography{Bibliography_base}


\begin{thebibliography}{22}


\ifx \showCODEN    \undefined \def \showCODEN     #1{\unskip}     \fi
\ifx \showISBNx    \undefined \def \showISBNx     #1{\unskip}     \fi
\ifx \showISBNxiii \undefined \def \showISBNxiii  #1{\unskip}     \fi
\ifx \showISSN     \undefined \def \showISSN      #1{\unskip}     \fi
\ifx \showLCCN     \undefined \def \showLCCN      #1{\unskip}     \fi
\ifx \shownote     \undefined \def \shownote      #1{#1}          \fi
\ifx \showarticletitle \undefined \def \showarticletitle #1{#1}   \fi
\ifx \showURL      \undefined \def \showURL       {\relax}        \fi
\providecommand\bibfield[2]{#2}
\providecommand\bibinfo[2]{#2}
\providecommand\natexlab[1]{#1}
\providecommand\showeprint[2][]{arXiv:#2}

\bibitem[Aher et~al\mbox{.}(2023)]%
        {Aher2023_using}
\bibfield{author}{\bibinfo{person}{Gati Aher}, \bibinfo{person}{Rosa~I.
  Arriaga}, {and} \bibinfo{person}{Adam~Tauman Kalai}.}
  \bibinfo{year}{2023}\natexlab{}.
\newblock \bibinfo{title}{Using Large Language Models to Simulate Multiple
  Humans and Replicate Human Subject Studies}.
\newblock
\showeprint[arxiv]{2208.10264}~[cs.CL]


\bibitem[Axelrod(1984)]%
        {Axelrod1984_evolution}
\bibfield{author}{\bibinfo{person}{Robert Axelrod}.}
  \bibinfo{year}{1984}\natexlab{}.
\newblock \bibinfo{booktitle}{\emph{The Evolution of Cooperation}}.
\newblock \bibinfo{publisher}{Basic Books}, \bibinfo{address}{New York}.
\newblock


\bibitem[Axelrod and Hamilton(1981)]%
        {Axelrod1981_evolution}
\bibfield{author}{\bibinfo{person}{Robert Axelrod} {and}
  \bibinfo{person}{William~D. Hamilton}.} \bibinfo{year}{1981}\natexlab{}.
\newblock \showarticletitle{The Evolution of Cooperation}.
\newblock \bibinfo{journal}{\emph{Science}} \bibinfo{volume}{211},
  \bibinfo{number}{4489} (\bibinfo{year}{1981}), \bibinfo{pages}{1390--1396}.
\newblock


\bibitem[Brookins and DeBacker(2023)]%
        {Brookins2023_playing}
\bibfield{author}{\bibinfo{person}{Philip Brookins} {and}
  \bibinfo{person}{Jason~M. DeBacker}.} \bibinfo{year}{2023}\natexlab{}.
\newblock \bibinfo{title}{Playing Games with {GPT}: What Can We Learn about a
  Large Language Model from Canonical Strategic Games?}
\newblock
\showeprint[arxiv]{2305.10912}~[econ.GN]


\bibitem[Chen et~al\mbox{.}(2021)]%
        {Chen2021_humaneval}
\bibfield{author}{\bibinfo{person}{Mark Chen}, \bibinfo{person}{Jerry Tworek},
  \bibinfo{person}{Heewoo Jun}, \bibinfo{person}{Qiming Yuan},
  \bibinfo{person}{Henrique~Ponde de Oliveira~Pinto}, \bibinfo{person}{Jared
  Kaplan}, \bibinfo{person}{Harri Edwards}, \bibinfo{person}{Yuri Burda},
  \bibinfo{person}{Nicholas Joseph}, \bibinfo{person}{Greg Brockman},
  {et~al\mbox{.}}} \bibinfo{year}{2021}\natexlab{}.
\newblock \bibinfo{title}{Evaluating Large Language Models Trained on Code}.
\newblock
\showeprint[arxiv]{2107.03374}~[cs.LG]


\bibitem[Fan et~al\mbox{.}(2024)]%
        {Fan2024_rational}
\bibfield{author}{\bibinfo{person}{Caoyun Fan}, \bibinfo{person}{Jindou Chen},
  \bibinfo{person}{Yaohui Jin}, {and} \bibinfo{person}{Hao He}.}
  \bibinfo{year}{2024}\natexlab{}.
\newblock \showarticletitle{Can Large Language Models Serve as Rational Players
  in Game Theory: A Systematic Analysis}. In
  \bibinfo{booktitle}{\emph{Proceedings of the AAAI Conference on Artificial
  Intelligence}}, Vol.~\bibinfo{volume}{38}. \bibinfo{pages}{17960--17967}.
\newblock


\bibitem[Guo(2023)]%
        {Guo2023_gpt}
\bibfield{author}{\bibinfo{person}{Fulin Guo}.}
  \bibinfo{year}{2023}\natexlab{}.
\newblock \bibinfo{title}{{GPT} Agents in Game Theory Experiments}.
\newblock
\showeprint[arxiv]{2305.05516}~[econ.GN]


\bibitem[Hendrycks et~al\mbox{.}(2021)]%
        {Hendrycks2021_mmlu}
\bibfield{author}{\bibinfo{person}{Dan Hendrycks}, \bibinfo{person}{Collin
  Burns}, \bibinfo{person}{Steven Basart}, \bibinfo{person}{Andy Zou},
  \bibinfo{person}{Mantas Mazeika}, \bibinfo{person}{Dawn Song}, {and}
  \bibinfo{person}{Jacob Steinhardt}.} \bibinfo{year}{2021}\natexlab{}.
\newblock \showarticletitle{Measuring Massive Multitask Language
  Understanding}.
\newblock \bibinfo{journal}{\emph{International Conference on Learning
  Representations}} (\bibinfo{year}{2021}).
\newblock
\showeprint[arxiv]{2009.03300}


\bibitem[Knight et~al\mbox{.}(2016)]%
        {Knight2016_open}
\bibfield{author}{\bibinfo{person}{Vincent Knight}, \bibinfo{person}{Owen
  Campbell}, \bibinfo{person}{Marc Harper}, \bibinfo{person}{Karol Langner},
  \bibinfo{person}{James Campbell}, \bibinfo{person}{Thomas Campbell},
  \bibinfo{person}{Alex Carney}, \bibinfo{person}{Martin Chorley},
  \bibinfo{person}{Cameron Davidson-Pilon}, \bibinfo{person}{Kristian Glass},
  {et~al\mbox{.}}} \bibinfo{year}{2016}\natexlab{}.
\newblock \showarticletitle{An Open Framework for the Reproducible Study of the
  Iterated Prisoner's Dilemma}.
\newblock \bibinfo{journal}{\emph{Journal of Open Research Software}}
  \bibinfo{volume}{4}, \bibinfo{number}{1} (\bibinfo{year}{2016}),
  \bibinfo{pages}{e35}.
\newblock


\bibitem[Leibo et~al\mbox{.}(2017)]%
        {Leibo2017_multiagent}
\bibfield{author}{\bibinfo{person}{Joel~Z. Leibo}, \bibinfo{person}{Vinicius
  Zambaldi}, \bibinfo{person}{Marc Lanctot}, \bibinfo{person}{Janusz Marecki},
  {and} \bibinfo{person}{Thore Graepel}.} \bibinfo{year}{2017}\natexlab{}.
\newblock \showarticletitle{Multi-Agent Reinforcement Learning in Sequential
  Social Dilemmas}. In \bibinfo{booktitle}{\emph{Proceedings of the 16th
  International Conference on Autonomous Agents and Multi-Agent Systems
  (AAMAS)}}. \bibinfo{pages}{464--473}.
\newblock


\bibitem[Madaan et~al\mbox{.}(2023)]%
        {Madaan2023_self_refine}
\bibfield{author}{\bibinfo{person}{Aman Madaan}, \bibinfo{person}{Niket
  Tandon}, \bibinfo{person}{Prakhar Gupta}, \bibinfo{person}{Skyler Hallinan},
  \bibinfo{person}{Luyu Gao}, \bibinfo{person}{Sarah Wiegreffe},
  \bibinfo{person}{Uri Alon}, \bibinfo{person}{Nouha Dziri},
  \bibinfo{person}{Shrimai Prabhumoye}, \bibinfo{person}{Yiming Yang},
  {et~al\mbox{.}}} \bibinfo{year}{2023}\natexlab{}.
\newblock \showarticletitle{Self-Refine: Iterative Refinement with
  Self-Feedback}. In \bibinfo{booktitle}{\emph{Advances in Neural Information
  Processing Systems (NeurIPS)}}, Vol.~\bibinfo{volume}{36}.
\newblock


\bibitem[Moran(1958)]%
        {Moran1958_random}
\bibfield{author}{\bibinfo{person}{Patrick A.~P. Moran}.}
  \bibinfo{year}{1958}\natexlab{}.
\newblock \showarticletitle{Random Processes in Genetics}.
\newblock \bibinfo{journal}{\emph{Mathematical Proceedings of the Cambridge
  Philosophical Society}} \bibinfo{volume}{54}, \bibinfo{number}{1}
  (\bibinfo{year}{1958}), \bibinfo{pages}{60--71}.
\newblock


\bibitem[Nowak(2006)]%
        {Nowak2006_evolutionary}
\bibfield{author}{\bibinfo{person}{Martin~A. Nowak}.}
  \bibinfo{year}{2006}\natexlab{}.
\newblock \bibinfo{booktitle}{\emph{Evolutionary Dynamics: Exploring the
  Equations of Life}}.
\newblock \bibinfo{publisher}{Harvard University Press},
  \bibinfo{address}{Cambridge, MA}.
\newblock


\bibitem[Park et~al\mbox{.}(2023)]%
        {Park2023_generative}
\bibfield{author}{\bibinfo{person}{Joon~Sung Park}, \bibinfo{person}{Joseph~C.
  O'Brien}, \bibinfo{person}{Carrie~J. Cai}, \bibinfo{person}{Meredith~Ringel
  Morris}, \bibinfo{person}{Percy Liang}, {and} \bibinfo{person}{Michael~S.
  Bernstein}.} \bibinfo{year}{2023}\natexlab{}.
\newblock \bibinfo{title}{Generative Agents: Interactive Simulacra of Human
  Behavior}.
\newblock
\showeprint[arxiv]{2304.03442}~[cs.HC]


\bibitem[Payne and Alloui-Cros(2025)]%
        {Payne2025_strategic}
\bibfield{author}{\bibinfo{person}{Kenneth Payne} {and}
  \bibinfo{person}{Baptiste Alloui-Cros}.} \bibinfo{year}{2025}\natexlab{}.
\newblock \bibinfo{title}{Strategic Intelligence in Large Language Models:
  Evidence from Evolutionary Game Theory}.
\newblock
\showeprint[arxiv]{2507.02618}~[cs.AI]


\bibitem[Piatti et~al\mbox{.}(2024)]%
        {Piatti2024_cooperate}
\bibfield{author}{\bibinfo{person}{Giorgio Piatti}, \bibinfo{person}{Zhijing
  Jin}, \bibinfo{person}{Max Kleiman-Weiner}, \bibinfo{person}{Bernhard
  Sch{\"o}lkopf}, \bibinfo{person}{Mrinmaya Sachan}, {and}
  \bibinfo{person}{Rada Mihalcea}.} \bibinfo{year}{2024}\natexlab{}.
\newblock \showarticletitle{Cooperate or Collapse: Emergence of Sustainable
  Cooperation in a Society of {LLM} Agents}. In
  \bibinfo{booktitle}{\emph{Advances in Neural Information Processing Systems
  (NeurIPS 2024)}}.
\newblock
\showeprint[arxiv]{2404.16698}~[cs.AI]


\bibitem[Traulsen et~al\mbox{.}(2006)]%
        {Traulsen2006_fixation}
\bibfield{author}{\bibinfo{person}{Arne Traulsen}, \bibinfo{person}{Martin~A.
  Nowak}, {and} \bibinfo{person}{Jorge~M. Pacheco}.}
  \bibinfo{year}{2006}\natexlab{}.
\newblock \showarticletitle{Stochastic dynamics of invasion and fixation}.
\newblock \bibinfo{journal}{\emph{Physical Review E}}  \bibinfo{volume}{74}
  (\bibinfo{year}{2006}), \bibinfo{pages}{011909}.
\newblock
\href{https://doi.org/10.1103/PhysRevE.74.011909}{doi:\nolinkurl{10.1103/PhysRevE.74.011909}}


\bibitem[Vallinder and Hughes(2024)]%
        {Vallinder2024_cultural}
\bibfield{author}{\bibinfo{person}{Aron Vallinder} {and}
  \bibinfo{person}{Edward Hughes}.} \bibinfo{year}{2024}\natexlab{}.
\newblock \bibinfo{title}{Cultural Evolution of Cooperation among {LLM}
  Agents}.
\newblock
\showeprint[arxiv]{2412.10270}~[cs.MA]
\newblock
\shownote{Extended Abstract at AAMAS 2025}.


\bibitem[Wang et~al\mbox{.}(2024)]%
        {Wang2024_survey}
\bibfield{author}{\bibinfo{person}{Lei Wang}, \bibinfo{person}{Chen Ma},
  \bibinfo{person}{Xueyang Feng}, \bibinfo{person}{Zeyu Zhang},
  \bibinfo{person}{Hao Yang}, \bibinfo{person}{Jingsen Zhang},
  \bibinfo{person}{Zhiyuan Chen}, \bibinfo{person}{Jiakai Tang},
  \bibinfo{person}{Xu Chen}, \bibinfo{person}{Yankai Lin}, {et~al\mbox{.}}}
  \bibinfo{year}{2024}\natexlab{}.
\newblock \showarticletitle{A survey on large language model based autonomous
  agents}.
\newblock \bibinfo{journal}{\emph{Frontiers of Computer Science}}
  \bibinfo{volume}{18}, \bibinfo{number}{6} (\bibinfo{year}{2024}),
  \bibinfo{pages}{186345}.
\newblock


\bibitem[Willis et~al\mbox{.}(2025)]%
        {Willis2025_llm_ipd}
\bibfield{author}{\bibinfo{person}{George Willis}, \bibinfo{person}{Yali Du},
  \bibinfo{person}{Joel~Z. Leibo}, {and} \bibinfo{person}{Michael Luck}.}
  \bibinfo{year}{2025}\natexlab{}.
\newblock \bibinfo{title}{Do {LLM} Agents Cooperate or Defect? Evolutionary
  Dynamics in Multi-Agent Systems}.
\newblock
\showeprint[arxiv]{2501.16173}~[cs.GT]


\bibitem[Wu and Axelrod(1995)]%
        {WuAxelrod1995_noise}
\bibfield{author}{\bibinfo{person}{Jianzhong Wu} {and} \bibinfo{person}{Robert
  Axelrod}.} \bibinfo{year}{1995}\natexlab{}.
\newblock \showarticletitle{How to Cope with Noise in the Iterated Prisoner's
  Dilemma}.
\newblock \bibinfo{journal}{\emph{Journal of Conflict Resolution}}
  \bibinfo{volume}{39}, \bibinfo{number}{1} (\bibinfo{year}{1995}),
  \bibinfo{pages}{183--189}.
\newblock


\bibitem[Yocum et~al\mbox{.}(2023)]%
        {Yocum2023_mitigating}
\bibfield{author}{\bibinfo{person}{Julian Yocum}, \bibinfo{person}{Phillip
  Christoffersen}, \bibinfo{person}{Mehul Damani}, \bibinfo{person}{Justin
  Svegliato}, \bibinfo{person}{Dylan Hadfield-Menell}, {and}
  \bibinfo{person}{Stuart Russell}.} \bibinfo{year}{2023}\natexlab{}.
\newblock \showarticletitle{Mitigating Generative Agent Social Dilemmas}. In
  \bibinfo{booktitle}{\emph{Foundation Models for Decision Making Workshop,
  NeurIPS}}.
\newblock


\end{thebibliography}

\end{document}